\documentclass[conference]{IEEEtran}
\IEEEoverridecommandlockouts

\usepackage[T1]{fontenc}
\usepackage{cite}
\usepackage{amsmath,amssymb,amsfonts}
\usepackage{algorithmic}
\usepackage{graphicx}
\usepackage{textcomp}
\usepackage{xcolor}
\usepackage{booktabs}
\usepackage{hyperref}

\usepackage{enumitem}
\setlist[itemize]{leftmargin=3mm}
\usepackage[english]{babel}
\usepackage[utf8]{inputenc}
\usepackage{comment}

\begin{document}

\title{Don't Get Hijacked: Prevalence, Mitigation, and Impact of Non-Secure DNS Dynamic Updates}

\author{\IEEEauthorblockN{Yevheniya Nosyk\IEEEauthorrefmark{1}, Maciej Korczy\'nski\IEEEauthorrefmark{1},Carlos H. Gañán\IEEEauthorrefmark{3}, Michał Król\IEEEauthorrefmark{4}, Qasim Lone\IEEEauthorrefmark{5}, Andrzej Duda\IEEEauthorrefmark{1}}

\IEEEauthorblockA{\IEEEauthorrefmark{1}Univ. Grenoble Alpes, CNRS, Grenoble INP, LIG, France   \\ Email: firstname.lastname@univ-grenoble-alpes.fr \\
\IEEEauthorrefmark{3}TU Delft, The Netherlands \\
\IEEEauthorrefmark{4}City, University of London, The United Kingdom  \\
\IEEEauthorrefmark{5}RIPE NCC, The Netherlands
}
}

\maketitle

\begin{abstract} 
DNS dynamic updates represent an inherently vulnerable mechanism deliberately granting the potential for any host to dynamically modify DNS zone files. Consequently, this feature exposes domains to various security risks such as domain hijacking, compromise of domain control validation, and man-in-the-middle attacks. Originally devised without the implementation of authentication mechanisms, non-secure DNS updates were widely adopted in DNS software, subsequently leaving domains susceptible to a novel form of attack termed zone poisoning. In order to gauge the extent of this issue, our analysis encompassed over 353 million domain names, revealing the presence of 381,965 domains that openly accepted unsolicited DNS updates. We then undertook a comprehensive three-phase campaign involving the notification of Computer Security Incident Response Teams (CSIRTs). Following extensive discussions spanning six months, we observed substantial remediation, with nearly 54\% of nameservers and 98\% of vulnerable domains addressing the issue. This outcome serves as evidence that engaging with CSIRTs can prove to be an effective approach for reporting security vulnerabilities. Moreover, our notifications had a lasting impact, as evidenced by the sustained low prevalence of vulnerable domains. 

\end{abstract}


\section{Introduction}

In the early stages of the Internet, hosts were primarily identified by their IP addresses, which were difficult to remember. To alleviate this issue, the Stanford Research Institute introduced the static \texttt{HOSTS.TXT} file, which facilitated the mapping of host names to IP addresses. As the number of network-connected devices rapidly increased, a more scalable solution was required. The Domain Name System (DNS), standardized in 1987~\cite{rfc1034,rfc1035} met the requirement.

The early DNS was relatively static. Domain owners occasionally updated local copies of zone files, but the whole process was not automated and could not scale. However, with the emergence of the Dynamic Host Configuration Protocol (DHCP)~\cite{rfc2131}, it became essential to promptly assign domain names to dynamically added hosts. In 1997, the IETF published a new RFC 2136~\cite{rfc2136} proposed standard called ``Dynamic Updates in the Domain Name System (DNS UPDATE).'' This new mechanism allowed dynamically updating the content of zone files of authoritative nameservers. Notably, the authors stated that unless coupled with some external security mechanism, any host on the Internet would be able to update external zone files by sending a single UDP packet. As DNS was becoming increasingly ubiquitous and thus an attractive attack target, such protocol extensions became particularly dangerous for domain owners.

DNS has long been an attractive target for attackers. Domain shadowing~\cite{shadowing}, cache poisoning attacks~\cite{cachepoisoning}, and other forms of DNS manipulation~\cite{global_manipulation, Intercept} may remain unnoticeable for a while. Non-secure DNS updates make such attacks trivial---they eliminate the need to steal any credentials and allow modifying target zone files directly. Contrary to cache poisoning attacks that affect individual recursive resolvers, modified zone files are globally accessible. Adopting the terminology of the previous work~\cite{zonepoisoning}, we refer to the attacks exploiting non-secure DNS dynamic updates as \textit{zone poisoning}.

While zone poisoning attacks can have devastating consequences for domain owners, they received very little attention. To the best of our knowledge, only one paper enumerated domain names vulnerable to zone poisoning in the wild. In 2016, researchers scanned a sample of 2.9M randomly chosen domains and 1M domains from the Alexa popularity list. Although that study gives the initial glance at the prevalence of the vulnerability in the wild, the coverage is 
low, and previous attempts to mitigate the problem did not result in substantial remediation~\cite{cetin2017make}. In this paper, we extend the work of Korczy\'nski et al.~\cite{zonepoisoning} and present the following contributions:

\begin{enumerate}

    \item \textbf{We define an extensive zone poisoning attack taxonomy.} We present five classes of attacks that can be performed on domains supporting non-secure DNS dynamic updates. We experimentally verify those attacks in a controlled test environment.
    
    \item \textbf{We scan more than 353M domain names and 3.6M nameservers---two orders of magnitude more than previous work.} We test whether they accept non-secure DNS dynamic updates from arbitrary clients and identify 200 times more vulnerable domains (382K) and 5 times more (5.6K) vulnerable nameservers. We analyze their distribution across autonomous systems and countries.
    
    \item \textbf{We lead a large-scale notification campaign to fix vulnerable resources.} We send carefully-crafted notification emails containing various nudges to CSIRTs and actively engage in discussion with them. We show that contacting CSIRTs can be an efficient way to report vulnerabilities and have them fixed (contrary to the common disbelief). As a result, almost 54\% of nameservers and 98\% of vulnerable domain names were remediated. 

    \item \textbf{After carefully analyzing the attack, we followed a responsible disclosure procedure.} At the time of writing, two CVEs 
    have been reserved 
    for the impacted DNS vendor software: Knot DNS and Simple DNS Plus.
    
\end{enumerate}

\section{Background\label{sec:background}}

This section introduces the necessary background on dynamic updates in DNS, the associated security risks, and implementations in popular DNS software.

\subsection{Dynamic Updates in DNS}

Dynamic Updates in DNS were introduced in the proposed standard RFC 2136~\cite{rfc2136} back in 1997 to address the need to update the content of DNS zones dynamically. They allowed efficient updating, adding, or deleting any type of resource record (RR), e.g., \texttt{A}, \texttt{AAAA}, \texttt{NS}, etc. 

If the authoritative nameserver supports dynamic updates, it inspects the packet to identify whether all update prerequisites (if any) are met and whether the client is allowed to request such an update. Note that unless restricted by the nameserver, anyone who knows the zone name (e.g., \texttt{example.com}) and the nameserver (e.g., \texttt{ns1.example.com}) is capable of updating the zone content by sending a single UDP datagram.

\subsection{Security Considerations}

Dynamic updates raise a vital concern---if configured insecurely, they will be accepted from any host on the Internet. As acknowledged in RFC 2136, non-secure updates constitute ``a serious increase in vulnerability from the current technology.'' The subsequent RFC 2137~\cite{rfc2137} and RFC 3007~\cite{rfc3007} proposed using cryptographic keys to generate signatures covering update requests. This would only allow authorized clients to perform updates but adds more complexity related to key management. As an alternative, a lightweight security mechanism based on shared secret keys and one-way hashing was introduced three years later in RFC 2845 (TSIG: Secret Key Transaction Authentication for DNS)~\cite{rfc2845}. 

If none of the cryptographic security mechanisms is implemented, an authoritative nameserver is expected only to accept the dynamic updates from a statically preconfigured set of IP addresses. The address match lists should be as restrictive as possible and limited to, for example, an IP address of a DHCP server. Nevertheless, such a mechanism is still insecure because the adversary may guess the IP addresses from the list and send requests with spoofed source IPs. Such access control lists could have been effective if dynamic updates were using the TCP protocol for transport. However, Paul Vixie--the editor of RFC 2136--explained to us that the working group was considering proposing TCP rather than UDP, but that would involve opening the port TCP 53 in firewalls, which seemed to be problematic at that time. Therefore, they decided to release the document as a \textit{proposed standard} instead of an \textit{Internet standard} at least until the security issues raised in the specification are not addressed. 

\subsection{Implementation of Dynamic Updates}

We analyzed five popular implementations of DNS authoritative nameserver software and verified whether they support dynamic updates and what default settings are offered to clients as of April 2023:


\begin{itemize}
    \item \textbf{BIND 9.18.4:} by default, dynamic updates are disabled and must be explicitly configured using either \texttt{allow-update} or \texttt{update-policy} options. The first statement (\texttt{allow-update}) is a simple access control list (ACL) that grants permission to update the zone for any address matching the list. This option, however, raises several security issues. First, an administrator may use a built-in \texttt{any} argument to accept updates from all IP ranges. Second, even if the ACL is restricted to individual IP addresses, an attacker may send \texttt{UPDATE} requests with spoofed source IP addresses from the ACL. The official BIND9 manual~\cite{bind9manual} strongly recommends avoiding IP-based ACLs and specifying \texttt{TSIG} key names instead. The second statement (\texttt{update-policy}) is restrictive as it only allows \texttt{TSIG}-based access lists and does not let one specify a range of IP addresses. 

    \item \textbf{Windows Server 2022:} distinguishes between two types of zones: standard primary and directory-integrated. The latter support either ``secure only'' updates using extended \texttt{TSIG} or ``nonsecure and secure'' updates--a zone type that accepts updates from any client. Standard primary zone configuration supports \textit{only} ``nonsecure and secure''  updates. Microsoft is aware of the vulnerability and informs the users willing to set up a non-secure implementation that ``allowing nonsecure dynamic updates is a significant security vulnerability because updates can be accepted from untrusted sources.'' 

    \item \textbf{PowerDNS 4.6.2:} dynamic updates are implemented but disabled by default. They are explicitly activated in the main configuration file with the \texttt{dnsupdate=yes} statement. Updates are only accepted from IP address ranges defined under \texttt{allow-dnsupdate-from} (the default is \texttt{127.0.0.0/8}, but \texttt{0.0.0.0/0} would match any IPv4 address)~\cite{powerdns}. If the secondary server receives the update, it is automatically forwarded to the primary server (provided all the permissions are granted). More options can be configured per each individual zone, for example, the list of allowed addresses or \texttt{TSIG}. 

    \item \textbf{Knot DNS 3.1.9:} Dynamic updates are implemented but need to be explicitly enabled with the \texttt{acl} statement and added to each corresponding \texttt{zone} statement~\cite{knot}. By default, the \texttt{address} field in \texttt{acl} will match any source IP address unless a more precise range is provided. This is a serious security threat that allows any host on the Internet to send dynamic updates. Nevertheless, secure updates are also available and can be configured by adding a \texttt{key} statement and then referring to it inside the \texttt{acl}. Similarly to PowerDNS or BIND9, updates received by the secondary server are forwarded to the primary server.

    \item \textbf{Simple DNS Plus 9.1:} this software implements standard (non-secure) and \texttt{TSIG}-signed dynamic updates, both disabled by default~\cite{simplednsplus}. When configuring standard updates, the administrators can either accept them from any IP address or create an ACL of address ranges.

\end{itemize}

\section{Adversary Model\label{sec:adversary}}

In this section, we provide a taxonomy of attacks conducted by an adversary on a vulnerable authoritative nameserver. We begin by describing our experimental setup and proceed to outline the attack vectors tested in our laboratory environment. 

\subsection{Infrastructure Setup}

To define the adversary model and validate it in a controlled environment, we established our infrastructure consisting of two servers:

\begin{itemize}
    \item \textbf{The adversary:} We assume that our adversary has already conducted scans for vulnerable resources and possesses knowledge of the domain name and its authoritative nameserver. The adversary also requires the use of the standard DNS dynamic update utility \texttt{nsupdate} \cite{nsupdate} or a dedicated software for modifying the victim's zone. For more sophisticated attacks, the adversary may need to configure a mail, web, or DNS server.

    \item \textbf{The victim:} 
    We configure the \texttt{example.com} test zone and the \texttt{ns1.example.com} nameserver using BIND9 software. It is configured to accept non-secure dynamic updates from any host on the Internet, as governed by the \texttt{allow-update} option. Furthermore, we host a website using the Apache HTTP server software, deploy SSL/TLS certificates, and operate a mail server (\texttt{mail.example.com}) with the corresponding \texttt{MX} record defined in the zone file.

\end{itemize}

\subsection{Taxonomy of Attacks}

Below are five categories of zone poisoning attacks, each with its own trade-off between viability and stealthiness.

\subsubsection{Denial of Service (DoS) Attack} 

This category of attacks is relatively simple to execute, although they lack significant stealthiness, as the victim would promptly detect the unresponsive domain and take measures to rectify the configuration. 

\begin{itemize}
    \item  \textbf{Deletion of an \texttt{A}/\texttt{AAAA} record:} this removes the mapping between the domain name (\texttt{example.com}) and its corresponding IP address. Subdomains can be completely removed if managed by the same parent nameserver (\texttt{ns1.example.com}). If the victim operates different services, such as accounts, mail, FTP, or checkout, on separate subdomains, the adversary may selectively disable specific services, making detection less straightforward.
    
    \item \textbf{Deletion of an \texttt{MX} record:} similarly, the adversary can also delete the mail exchange record (and/or its glue record) that specifies the name of the mail server that accepts email messages on behalf of the domain name. It will not only disrupt the email service itself but also hinder any abuse (notification) messages sent to the victim. 

   \item \textbf{Deletion, addition, or modification of a \texttt{TXT} record:} \texttt{TXT} records are widely used for DNS-based service discovery, Domain-based Message Authentication, Domain Keys Identified Mail, DMARC, SPF, and more \cite{rfc6763,rfc6376,rfc7489,rfc7208}. 
   
   For example, an attacker can modify an existing \texttt{TXT} SPF record to \texttt{v=spf1 -all}, blocking all hosts from sending emails on behalf of \texttt{example.com}. While this attack is highly feasible, its stealthiness diminishes due to prompt detection of the manipulated \texttt{TXT} record. 
\end{itemize}

\subsubsection{Domain Hijacking Attack\label{hijacking}}

Modifies original DNS records so that the legitimate traffic is redirected to bogus servers under the attacker's control.

\begin{itemize}
    \item \textbf{Update of an \texttt{A/AAAA} record:} an adversary can update the victim's \texttt{A/AAAA} record and replace the legitimate IP address with the one under the attacker's control. As a result, all the client traffic (e.g., website visitors)  will be diverted from the domain owner to the adversary.
    
    \item \textbf{Update of an \texttt{MX} record:} if the victim operates a mail server, the adversary can access all email traffic associated with the victim's domain. They can achieve this by setting up a malicious mail server, such as \texttt{mail.malicious.com}, and modifying the \texttt{MX} record. If the vulnerable nameserver manages the domain, the adversary can update the \texttt{A/AAAA} record of \texttt{mail.example.com} with their own IP.

\end{itemize}

\subsubsection{Man-in-the-Middle (MITM) Attack} 

This advanced attack requires greater sophistication from the adversary, making it highly stealthy and difficult to detect. The adversary not only manipulates DNS records like in domain hijacking attacks but also intercepts and redirects the traffic to the victim's server.

\begin{itemize}
    \item \textbf{Update of an \texttt{A/AAAA} record:} the adversary may establish a proxy between the victim and its clients to either passively observe the traffic or create a malicious service to exploit the victim's customers and extract sensitive information like login credentials. In both cases, the adversary initiates the attack by modifying an \texttt{A/AAAA} record of a domain or subdomain.
    
    \item \textbf{Update of an \texttt{MX} record:} similarly to domain hijacking, the adversary must update the \texttt{MX} resource record to redirect all traffic to the malicious mail server. However, in this case, the mail traffic will be further forwarded to the intended recipients on behalf of the hijacked victim.
\end{itemize}

\subsubsection{Domain Shadowing Attack} 

Involves creating malicious subdomains for exploit kits, malware, phishing, and client information theft. These attacks are highly stealthy as subdomains leverage the trust of the parent domain. The adversary can create multiple subdomains effortlessly and rotate between them to avoid detection.

\begin{itemize}

    \item \textbf{Addition of an \texttt{A/AAAA} record:} create numerous malicious subdomains that direct to web servers hosting, for example, malware or phishing websites. This involves adding corresponding \texttt{A} or \texttt{AAAA} records to the zone file, such as assigning \texttt{1.2.3.4} to \texttt{paypal.account.example.com}. To enhance the attack's stealthiness and evade IP-based blocklisting, the adversary may employ the fast-flux technique, dynamically associating multiple subdomains with a wide range of IP addresses. 

    \item \textbf{Addition of an \texttt{NS} record:} Generating multiple subdomains and rotating IP addresses helps evade blocklisting services, but frequent changes in the zone file may alert the victim. 
    To enhance the stealthiness of the domain shadowing attack, the adversary can introduce a malicious delegation (\texttt{account.example.com}) and set up a nameserver (\texttt{ns1.account.example.com}). This allows the adversary to create subdomains (\texttt{paypal. account.example.com}) directly on the malicious nameserver, bypassing the victim's server. 
    
\end{itemize}

\subsubsection{Compromising Domain Control Validation\label{sec:domainvalidation}} 

Digital certificates establish domain authenticity, verified through Domain Control Validation (DCV) to prove ownership. However, domain hijacking and other techniques discussed can bypass this validation. These stealthy attacks involve temporary zone file modifications during certificate issuance.

\begin{itemize}
    
    \item \textbf{Update of an \texttt{A/AAAA} record:} HTTP-based validation requires uploading a file from the certification authority to a particular directory on the web server. The file must remain accessible over HTTP. For example, to obtain a certificate for \texttt{example.com}, a file must be uploaded accessible on \texttt{http://example.com/.well-known/pki} \texttt{-validation/filename.txt} to validate the ownership. If the adversary temporarily updates the web server's IP address (\texttt{A/AAAA} record), the certificate authority will look for a file on a malicious web server.
    
    \item \textbf{Addition of a \texttt{CNAME} record:} DNS-based validation may require domain owners to add a \texttt{CNAME} record to the zone file. The record contains a random string generated by the certificate authority, which requests the \texttt{CNAME} over DNS. If the adversary uploads such a record to the victim zone file, the domain ownership will be validated, and the attacker will obtain the digital certificate.
\end{itemize}

\subsection{Adversary Capabilities: Additional Insights}

This section explores the adversary's capabilities in zone poisoning attacks, specifically regarding the propagation of updates between primary and secondary nameservers and the vulnerability of IP-based access control lists to IP spoofing \cite{spoof}. These insights highlight the need for secure dynamic updates and TSIG-based access lists to mitigate the risks associated with these attack vectors.

\subsubsection{Propagation Between Primary and Secondary Nameservers}

A single DNS zone may be served by multiple nameservers (usually a primary and a secondary), so we further investigate whether unsolicited DNS updates would propagate between the two. We set up a primary and a secondary nameservers, but only enable dynamic updated at one of those at a time. In the first case, only the primary nameserver received an update packet. We confirmed that the newly added resource record was immediately propagated to the secondary nameserver via the DNS Incremental Zone Transfer Protocol (IXFR) mechanism. In the second scenario, the secondary nameserver accepted non-secure dynamic updates from arbitrary clients while the primary allowed updated from the secondary only. Upon sending the update packet to the secondary nameserver, it did not update its zone file directly but rather forwarded the request to the primary nameserver. The primary, in turn, updated its zone file and initiated a zone transfer to the secondary, thus adding a new resource record to both nameservers. 

\subsubsection{Dynamic Updates with Spoofed Source IPs}

In the aforementioned attack scenarios, we configured authoritative nameservers to freely accept non-secure updates. To prevent unsolicited zone changes, DNS administrators may create IP-based 
ACLs that would only allow authorized hosts (e.g., secondary nameservers, DHCP servers, machines from the same local network, etc.) to update zone files. Such configuration still remains highly insecure---as update packets are sent over UDP, an attacker can guess the authorized IP address and request a zone update on its behalf. We experimentally verified this attack scenario by configuring our nameserver to only accept updates from a particular IP address. We then sent an update packet with the spoofed source IP address from a different machine and confirmed that it was accepted by the nameserver. IP spoofing can, therefore, greatly increase the attack surface and target those nameservers that are seemingly secured and protected from arbitrary zone updates.
    
\section{Enumeration of Vulnerable Resources\label{sec:enumeration}}

This section identifies the domains and the corresponding authoritative nameservers vulnerable to zone poisoning. 

\begin{table}
    \scriptsize
    \setlength{\tabcolsep}{3pt}
    \caption{Tested and vulnerable resources identified during the global and subdomains scanning campaigns in February-March 2017 \label{tested-vulnerable}}
    \begin{tabular}{lcccc}
        \toprule
        & \multicolumn{2}{c}{\textbf{Global Scan}} & \multicolumn{2}{c}{\textbf{Subdomains Scan}} \\
        \cmidrule(lr){2-3}
        \cmidrule(lr){4-5}
        & All tested & Vulnerable & All tested & Vulnerable \\
        \midrule
        Domains & 353,870,510 & 381,965 (0.108\%) & 35,382,217 & 399 (0.0011\%)\\
        NS IPs & 3,855,615 & 5,575 (0.145\%) & 722,989 & 401 (0.0555\%)\\
        Domain--NS IPs & 5,032,117,394 & 679,930 (0.014\%) & 104,955,041 & 520 (0.0005\%)\\
        \bottomrule
    \end{tabular}
\end{table}

\subsection{DNS Datasets \label{sec:dataset}}

\subsubsection{Global Scan} 

Sending a dynamic update requires two pieces of information: the zone name and the nameserver's IP address. Both can be found in passive DNS datasets or queried actively. We aggregated seven passive DNS datasets: i) Farsight's DNSDB~\cite{dnsdb}, ii) Censys's Internet-Wide Scan Data Repository~\cite{scansio}, iii) \texttt{.com}, \texttt{.net} and \texttt{.name} zone files provided by Verisign \cite{verisign}, iv) \texttt{.nl} zone file under the contract with SIDN--the \texttt{.nl} ccTLD registry~ \cite{sidn}, v) \texttt{AXFR} transfers of \texttt{.se} and \texttt{.nu} ccTLD \cite{se}, vi) \texttt{.us} ccTLD, \texttt{.biz}, \texttt{.org}, \texttt{.asia}, \texttt{.info}, \texttt{.mobi}, \texttt{.post} and \texttt{.tel} legacy gTLDs and 1230 new gTLDs made available by ICANN through the Centralized Zone Data Service \cite{icann-czds}, and vii) Alexa Top 1M \cite{alexa}.

From each dataset, we extracted all the second-level and upper-level domain names (if registered under public suffixes~\cite{publicsuffix}, e.g., \texttt{example.co.uk}). Next, we checked whether passive datasets contained nameserver records (\texttt{NS}) and the corresponding glue records (IPv4 addresses). If not available, we actively queried the missing data. As shown in Table~\ref{tested-vulnerable}, we gathered more than 353M unique domain names and 3.8M unique IPv4 addresses of nameservers, which render more than 5B domain name--nameserver pairs for scanning.

\subsubsection{Subdomains Scan} 

In DNS, a registered domain may contain multiple subdomains, possibly served by different nameservers. In the case of such delegations, parent and child nameservers can have different DNS dynamic update policies. We extract all the domains with three or more labels (depending on the public suffix's length) from Farsight's~\cite{dnsdb} passive query traces. We later compare whether parent and child (subdomain) authoritative nameservers have consistent configurations for DNS dynamic updates. We additionally queried all the missing data (i.e., \texttt{NS} records and/or their corresponding IPv4 addresses). From 35M domain names and 722K nameservers, we created a scanning input list with approximately 105M entries (see Table~\ref{tested-vulnerable}). 

\subsection{Scanning Methodology}

We developed an efficient scanner capable of sending DNS update packets at scale. Each update attempts to insert a new \texttt{A} resource record to the zone file. Specifically, we add a new subdomain in format \texttt{researchstudyzp.example.com}, where \texttt{example.com} is the tested zone's name.  The IP address is the one of our web server. It hosts a web page describing who we are, why we send DNS updates, how to correctly configure the server, and how to contact us. 

To undoubtedly confirm a vulnerability, we perform an active DNS lookup to resolve the subdomain to our web server's IP address. If this succeeds, then the vulnerability is present.  Finally, we remove the test DNS record by sending a delete request and then try to resolve it again to confirm the removal. We designed our scanning methodology so that only one UDP packet would be enough to test whether nameservers are vulnerable to the zone poisoning attack. We further discuss the ethical aspects of this study in Section~\ref{sec:ethical}.

\subsection{Scan Results}

We performed the global scan of domains and nameservers vulnerable to zone poisoning attack during four weeks in February-March 2017. Table~\ref{tested-vulnerable} presents the proportion of vulnerable resources found in each category (domain names, nameserver IP addresses, and domain-nameserver IP address pairs)---orders of magnitude more than the previous work~\cite{zonepoisoning}. While the ratio remains low (less than 1\% for each category), it translates into considerable absolute numbers. Overall, almost 382K domain names were found vulnerable to zone poisoning attacks. Given that those were often reachable over multiple vulnerable nameservers, 680K combinations of domain names and nameservers accepted arbitrary update requests.

To understand the population of vulnerable domains, we categorized them using Webshrinker\cite{webshrinker}---a tool that uses artificial intelligence to automatically classify domains under the IAB taxonomy~\cite{aerserv}. Such services are usually most effective when applied to domains that host websites, which is not necessarily the case in our input list. Nevertheless, we categorized 1.5\% of vulnerable domains. The ``Non-Standard Content'' is the most popular category representing message boards, content servers, or adult content (1089 domain names). The second most popular category is ``Business'', containing almost 800 domains, including banks and financial institutions. There is also a significant number of domains belonging to governmental (359), educational (483), and healthcare institutions (302). Therefore, some of the vulnerable domains represent critical services that are generally expected to be well-secured.

\begin{figure}[t]
    \centering
    \includegraphics[width=\columnwidth]{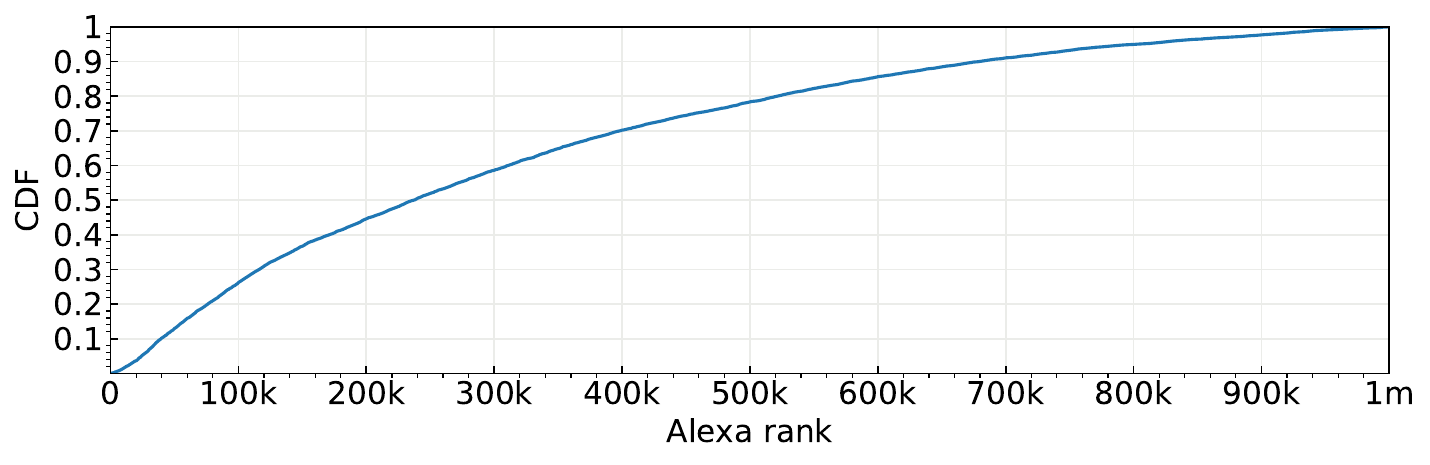}
    \caption{The distribution of 5,964 vulnerable domains in the Alexa domain popularity list.}
    \label{fig:alexa}
\end{figure}

To further access the popularity of vulnerable domain names, we aggregate the Alexa 1M~\cite{alexa} top website lists in 2017. We found 5,964 vulnerable domains in the ranking (the most popular domain reaching the 244\textsuperscript{th}  place on the list) and plotted the distribution of ranks in Figure~\ref{fig:alexa}. Overall, vulnerable domains are evenly distributed across the popularity~list.

We followed the global scan with a subdomains scan. The results are presented in Table~\ref{tested-vulnerable}. The ratio of identified vulnerable resources is up to 3 orders of magnitude lower than for the global scan. This comes from the nature of our input list. Lengthy multiple-level domains are likely to be disposable (i.e., generated for one-time use)~\cite{chen2014dns}. As a result, only 399 domains were vulnerable out of more than 35M and only 401 nameservers accepted non-secure updates for subdomains. Furthermore, we aggregated all the vulnerable subdomains by their corresponding second-level domains. Only 14 out of 236 aggregated second-level domains were vulnerable to zone poisoning. This finding highlights that subdomain servers may be vulnerable to our attack even when the delegating second-level domain servers are properly~configured. 

\begin{figure*}[t]
    \centering
    \includegraphics[width=\textwidth]{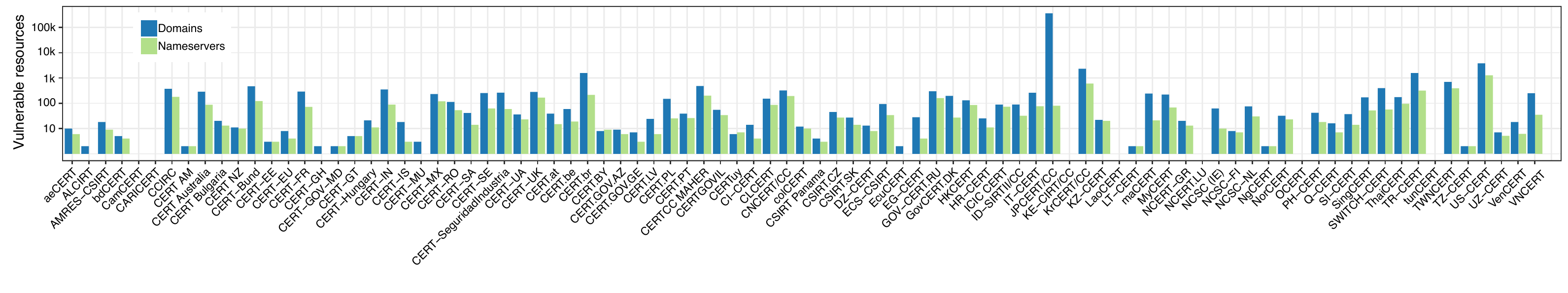}
    \caption{The number of vulnerable resources (domains and nameservers) per national CSIRT}
    \label{fig:certs}
\end{figure*}

\subsection{Unveiling Zone Poisoning Attacks \label{sec:unvealing}}

Zone poisoning attacks were previously unreported, prompting our investigation into their existence in the wild. Following the methodology outlined by Korczy\'nski et al.~\cite{zonepoisoning}, we utilized one year of passive DNS data provided by Farsight Security (2017). Our analysis focused on extracting queries related to subdomains of domains susceptible to zone poisoning, resulting in a total of 703K entries. 

Our hypothesis is that if attackers were to exploit regular domains accepting non-secure updates from arbitrary clients, they might engage in domain shadowing, creating subdomains involved in malicious activities. These subdomains may then be reported and added to URL or domain blocklists. To validate our hypothesis, we compared our domain list with APWG~\cite{apwg} and Phishtank~\cite{phishtank} feeds from the same period.  Remarkably, only three subdomains of vulnerable domains appeared in the Phishtank list. However, we found no evidence of their involvement in any abusive activities.

\subsection{Descriptive Statistics of Vulnerable Resources~\label{sec:statistics}}

Given the high absolute number of vulnerable domains and nameservers, it is crucial to aggregate them so that we can further notify the affected entities more efficiently.  After the introduction of GDPR \cite{temp}, WHOIS databases provide very limited to no contact information about domain owners. Therefore, we explore the distribution of vulnerable domains and nameservers across autonomous systems (ASes) and Computer Security Incident Response Teams (CSIRTs). 

\subsubsection{Per-AS statistics} Table~\ref{tested-vulnerable} shows that only 5.6K DNS nameservers are authoritative for the 382K vulnerable domains. On average, there are 121 domains per nameserver IP, but we identified three nameservers responsible for as many as 87\% of all the vulnerable domain names. The AS distribution of vulnerable nameservers is diverse as they originate from 1,682 ASes. We note, however, that while four ASes host more than 100 vulnerable nameservers each, they do not translate into a large number of vulnerable domains. On the contrary, a single AS from Japan hosts nameservers responsible for 95.4\% of vulnerable domains. As a result, while AS operators could potentially be our points of contact, it would require engaging with as many as 1.6K different entities.

\begin{figure}[t]
    \centering
    \begin{minipage}[b]{0.48\columnwidth}
        \includegraphics[width=\textwidth]{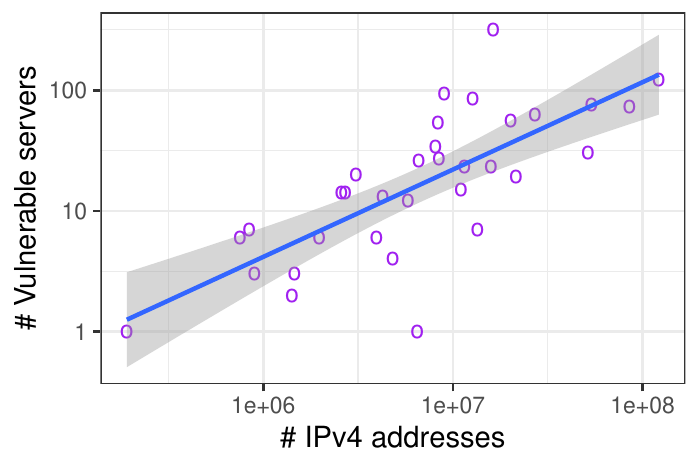}
    \end{minipage}
    \hfill
    \begin{minipage}[b]{0.48\columnwidth}
        \centering
        \includegraphics[width=\textwidth]{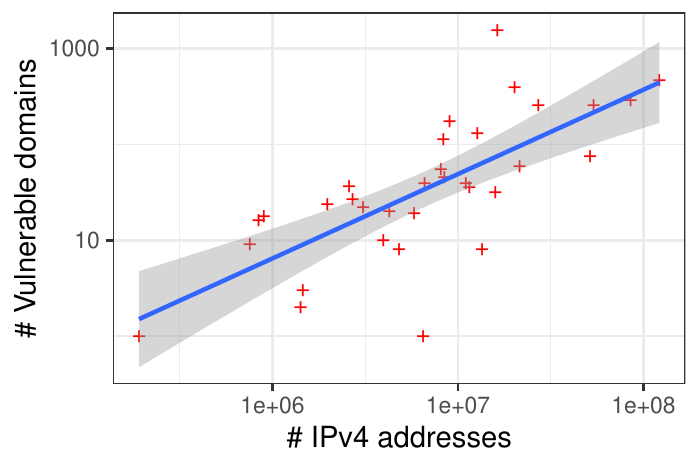}
    \end{minipage}
    \caption{The distribution of vulnerable resources (domains and nameservers) per national CSIRTs with respect to their size (number of IPv4 addresses under their jurisdiction).}
   \label{fig:size_cert}
\end{figure}

\subsubsection{Per-CSIRT statistics} Vulnerable nameservers are spread across 121 countries, with five of them (USA, South Korea, Taiwan, Turkey, and Iran) hosting almost half of all the vulnerable nameservers worldwide. Reporting vulnerabilities at the country level can be accomplished via CSIRTs. For each country hosting vulnerable nameservers and domains, we identified corresponding national CSIRTs (note that there can be multiple entities per country). We then computed the number of vulnerable domains and nameservers under the jurisdiction of each CSIRTs and plotted the result in Figure~\ref{fig:certs}. Following the country-level pattern, the Japanese CSIRT is responsible for 2-orders of magnitude more vulnerable domains than any other CSIRTs. Aside from the JP-CERT, the distribution of vulnerable resources across national CSIRTs follows a log-normal distribution. Thus, while some CSIRTs have only one vulnerable resource, more than 50\% of CSIRTs are responsible for hundreds of vulnerable domain names and nameservers. 

To further investigate this log-normal distribution, Figure~\ref{fig:size_cert} plots: a) vulnerable authoritative nameserver IPs and b) vulnerable domain IPs (\texttt{A} records) to the IPv4 address space under the jurisdiction of each CSIRT. As expected, larger CSIRTs accumulate a higher number of vulnerable resources. Nevertheless, there are a few outliers. The Tunisian CSIRTs had only 1 vulnerable resource for a total of 6M IPv4 addresses. On the contrary, the Malaysian CSIRTs of similar size had more than 200 vulnerable resources.

\section{Notifications\label{sec:notifications}}

Non-secure dynamic updates pose a significant threat to domain owners. To motivate remediation action, we conducted a series of notification campaigns. 

\begin{table}
    \caption{The summary of notification campaigns \label{tab:notif_campaign}}
    \scriptsize
    \begin{tabular}{lcccc}
        \toprule
        & \textbf{Date} & \textbf{Notified} & \textbf{Unreachable} & \textbf{Replies}\\
        \midrule
        Phase 0 & 2017-05-01 & 1 & - & 1 manual\\
        \midrule
        Phase 1 & 2017-09-06 & 44 & 2 & 7 automatic / 16 manual \\
        & 2017-09-28 & 35 & 2 & 6 automatic / 13 manual \\
        & 2017-10-19 & 27 & 2 & 4 automatic / 7 manual \\
        \midrule
        Phase 2 & 2018-02-14 & 168 & 5 & 14 automatic / 40 manual \\
        & 2018-02-28 & 167 & 5 & 12 automatic / 24 manual \\
        & 2018-03-16 & 162 & 5 & 12 automatic / 24 manual \\
        & 2018-04-12 & 76 & 5 & 7 automatic / 7 manual \\
        \bottomrule
    \end{tabular}
\end{table}

\begin{table*}
    \caption{Remediation effects by the end of Phase 1 \label{tab:phase1_summary}}
    \scriptsize
    \begin{tabular}{lcccc}
        \toprule
        & \multicolumn{2}{c}{\textbf{TF-CSIRTs (N = 44)}} & \multicolumn{2}{c}{\textbf{Other CSIRTs (N = 36)}}\\
        \cmidrule(lr){2-3}
        \cmidrule(lr){4-5}
        & \textbf{Servers (Remediated/Vulnerable)} & \textbf{Domain (Remediated/Vulnerable)} & \textbf{Servers (Remediated/Vulnerable)} & \textbf{Domains (Remediated/Vulnerable)} \\
        \midrule
        \textbf{N (\%)} & 328 (20.4\%) / 1280 (79.6\%) & 737 (14.24\%) / 4439 (85.76\%) & 658 (19.17\%) / 2775 (80.83\%) & 2173 (22.29\%) / 7574 (77.71\%)\\
        \textbf{max} & 62 / 235 & 108 / 1231 & 220 / 911 & 847 / 2384 \\
        \textbf{median} & 3.0 / 14.5 & 4.0 / 29.5 & 2.5 / 13.0 & 4.5 / 34.0 \\
        \textbf{mean (sd)} & 7.45 $\pm$ 11.71 / 29.09 $\pm$ 43.03 & 16.75 $\pm$ 27.41 / 100.89 $\pm$ 195.19 & 18.28 $\pm$ 39.10 / 77.08 $\pm$ 172.45 & 60.36 $\pm$ 151.84 / 210.39 $\pm$ 494.86 \\
        \bottomrule
    \end{tabular}
\end{table*}

\subsection{Notification Methodology}

Choosing the right notification recipient ensures that the message will be understood and appropriate actions will be taken. We notified national CSIRTs---organizations responsible for reacting to cyber threats and ensuring the hygiene of networks under their jurisdiction. For each vulnerable resource, we identified the responsible entity and extracted the contact email address using CERT.at's database~\cite{cert}.

The subject line of our emails informs on the number of resources still vulnerable to zone poisoning. Specifically, it states ``\texttt{XX} domain(s) still vulnerable to zone poisoning, \texttt{YY} nameservers fixed'', where \texttt{XX} represents the number of vulnerable domains at the time of the notification and \texttt{YY} stands for the number of nameservers fixed since the global scan. It serves as a reminder nudge~\cite{nudging} to recall the recipients what is still to be fixed. 

The body of the email contains four sections: i) a high-level description of the problem of non-secure dynamic updates, ii) the list of vulnerable nameservers and domain names, iii) names of organizations managing vulnerable resources, and iv) the list of necessary steps to fix the insecure configuration together with a pointer to a more extensive guide.

Table~\ref{tab:notif_campaign} summarizes the three phases of our notification campaign: \textit{Phase 0} targeting the Japanese CSIRT, \textit{Phase 1} targeting CSIRT members of the so-called ``Trusted Introducer'' community, and \textit{Phase 2} targeting national and governmental CSIRTs. In total, we contacted 200 entities over six months. 

\begin{figure}[t]
    \centering
    \includegraphics[width=0.85\columnwidth]{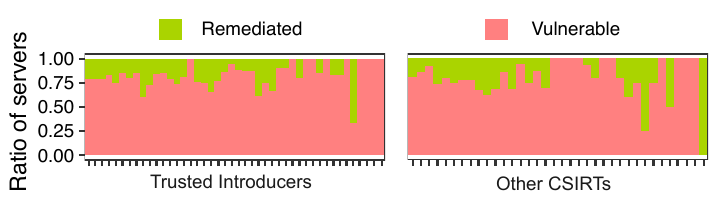}
    \caption{Remediation rates of the DNS nameservers during Phase 1 under two groups of CSIRTs - Trusted Introducers and Others}
    \vspace{-0.2cm}
    \label{fig:remediation_phase1}
\end{figure}

\subsection{Phase 0: JP-CERT}

Given the high concentration of vulnerable resources in Japan, we contacted the JP-CERT before launching the large-scale notification campaign. While 29 nameservers were fixed and 92\% of Japanese domains were not vulnerable anymore, the JP-CERT could not obtain more details from the affected network operators. To raise awareness in Japan and support the remediation action, JP-CERT wrote an article regarding zone poisoning and countermeasures.

After Phase 0, there remained more than 5K servers worldwide that accepted non-secure DNS updates from arbitrary clients, exposing more than 43K domains at risk of being exploited. We thus proceeded with the first phase of the notification campaign.

\subsection{Phase 1: The Trusted Introducer Service}

The Trusted Introducer Service (Task Force-CSIRT)\cite{tf-csirt} is a European CSIRT community formed in 2000. Its goal is to meet shared needs and construct a service architecture that provides critical assistance for all security and incident response teams. From September 7 to October 19, 2017, we sent emails to 44 teams that had vulnerable nameservers under their jurisdiction. While two CSIRTs were unreachable at indicated email addresses, around half of the remaining ones acknowledged the reception of our notifications either manually or via the creation of automatic tickets. We provided further information on the vulnerability to 5\% of CSIRTs.

We then compare the remediation rates of the notified population against a random sample of 36 CSIRTs that were not part of the TF-CSIRT community at the time of the campaign. Table~\ref{tab:phase1_summary} presents the results at the end of Phase 1. Notified CSIRTs remediated 20.4\% of corresponding nameservers and 14.24\% of domains. We observe similar rates for the non-notified parties. This so-called ``natural remediation'' could be the result of i) transient misconfigurations that were detected independently, or ii) information received by the non-notified CSIRTs extraneous to our notification campaign.

Not all the CSIRTs contributed equally to the overall remediation. We tracked the population of vulnerable nameservers for three months and then computed the ratio of remediated machines to all those tested during the global scan. Figure~\ref{fig:remediation_phase1} shows that while some CSIRTs fixed the considerable part of vulnerable nameservers, the average remediation rate per CSIRT was 26.2\%. This highlights that CSIRTs are generally effective at remediating vulnerabilities but do not eliminate the problem completely.

\begin{figure}[t]
    \centering
    \includegraphics[width=0.8\columnwidth]{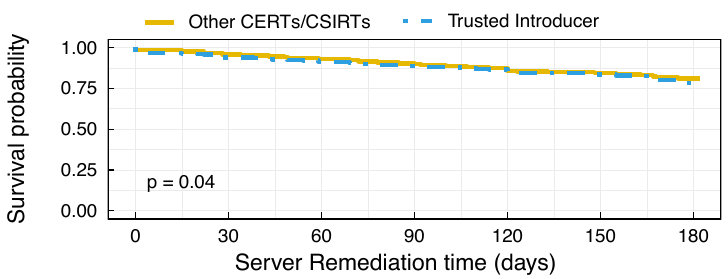}
    \caption{Survival rates of vulnerable nameservers during Phase 1.}
    \label{fig:km_serverspha1}
\end{figure}

\begin{figure}[t]
    \centering
    \includegraphics[width=0.8\columnwidth]{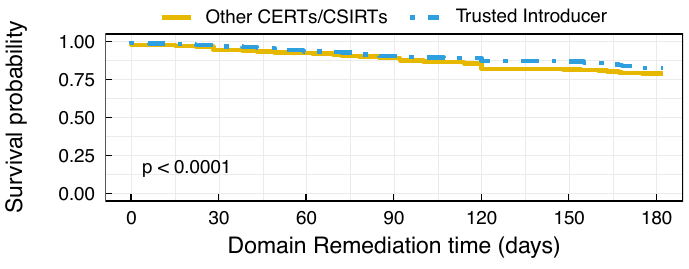}
    \caption{Survival rates of vulnerable domains during  Phase~1.}
    \label{fig:km_dompha1}
    \vspace{-0.2cm}
\end{figure}

We further compute the survival curves by using Kaplan-Meier estimates and interval censoring, defining the survival time as the time from the moment the notification was sent till the moment the resource was remediated. Figures~\ref{fig:km_serverspha1} and \ref{fig:km_dompha1} plot survival curves for vulnerable nameservers and domains, respectively. While their shape is similar, the Gehan-Breslow-Wilcoxon test indicates a significant difference between the curves. It suggests that our notifications were effective shortly after being sent and triggered better remediation rates compares to those CSIRTs that were not notified at all.

\begin{figure}[t]
    \centering
    \includegraphics[width=0.8\columnwidth]{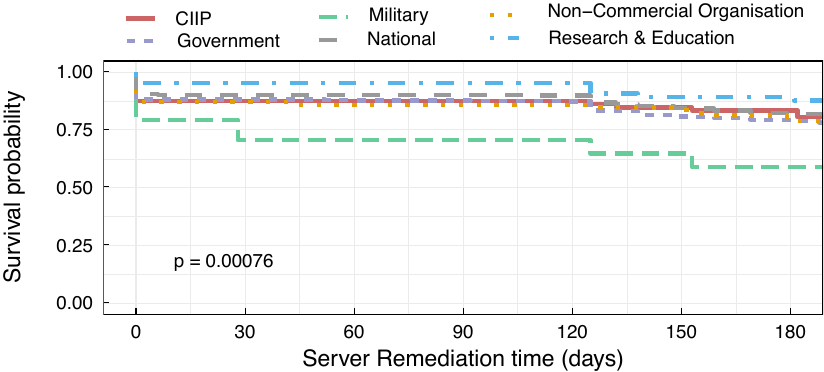}
    \caption{Server remediation rates of different CSIRT constituency types.}
    \vspace{-0.2cm}
    \label{fig:csirt_tyoe}    
\end{figure}

There exist different types of CSIRTs, e.g., governmental, national, military, research and education, non-commercial, and critical information infrastructure (CIIP). We wonder if the sector in which a CSIRT operates impacts the remediation rate. Figure~\ref{fig:csirt_tyoe} shows the survival curves for vulnerable nameservers with respect to the CSIRT type. We again refer to the Gehan-Breslow-Wilcoxon test to confirm that there is a statistically significant difference between the different types. Military CSIRTs are particularly fast and efficient in remediation, contrary to research and educational entities that exhibit the slowest and worst patching rates. 

\begin{figure}[t]
    \centering
    \includegraphics[width=0.8\columnwidth]{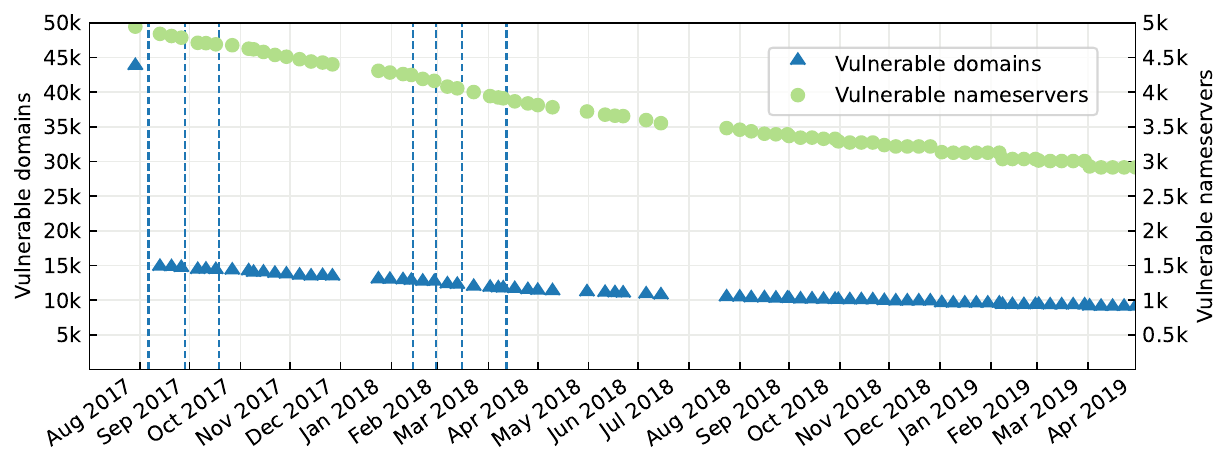}
    \caption{Number of vulnerable resources over time. The vertical dashed lines represent two notification campaigns - Phase 1 and Phase 2.}
    \vspace{-0.2cm}
    \label{fig:ts_notif}
\end{figure}

\subsection{Phase 2: National CSIRTs}

We conducted the second phase of notifications three months after Phase 1. This time, we went beyond the Trusted Introducer Service and expanded our campaign to all the national CSIRTs with vulnerable resources under their jurisdiction. We also modified the notification email and included a link to our website containing all the aggregated statistics about the number of vulnerable resources globally. This allowed all visitors to track their mitigation progress over time, but also to compare themselves against the rest of the CSIRTs. We hypothesized that this social norms nudge~\cite{nudging} would encourage CSIRTs to fix the vulnerability.

In total, we notified 168 national CSIRTs. Figure~\ref{fig:ts_notif} plots the number of vulnerable resources observed during Phase 1 and Phase 2 notification campaigns. The vertical dashed lines refer to the dates when the emails were sent---three days in 2017 for Phase 1 and four days in 2018 for Phase 2. This second notification campaign accelerated the remediation rates, although not all the CSIRTs were equally efficient. Around 2\% of national CSIRTs reached back to inform us that acting upon this vulnerability was not within their mandate. 

Overall, the three notification campaigns led to fixing 97.96\% of vulnerable domains and 53.59\% of vulnerable nameservers. These remediation rates are higher than the ones reported by previous studies and signal the need to involve CSIRTs in the remediation of vulnerabilities.  

\section{Long-Term Impact\label{sec:long-term}}

We now assess the lasting impact of our three notification campaigns on securing domains and nameservers from zone poisoning. Four years later, we perform active measurements to gauge the increase in newly vulnerable domain names and the persistence of previously identified vulnerable resources.


Our input dataset was collected in June 2022 and comprised the following sources: i) Tranco domain popularity list~\cite{tranco}, ii) Certificate Transparency logs from Calidog~\cite{calidog}, iii) passive DNS query traces from SIE Europe~\cite{sieeurope}, iv) 1150 legacy gTLD and new gTLD zone files provided by ICANN's CZDS~\cite{icann-czds}, v) AXFR transfers for \texttt{.se}, \texttt{.nu}, \texttt{.ch}, and \texttt{.li} zones. During this campaign, we aggregated all the domain lists and queried \texttt{SOA} records to only keep active domain names for further analysis. We then requested nameserver resource records (\texttt{NS}) and their corresponding IP addresses (\texttt{A} and \texttt{AAAA} records). Moreover, we extended our measurements to IPv6 address space. 

The data sets for both the 2017 and 2022 scans are comparable, including the vast majority of domain names under gTLDs (\texttt{.com}, \texttt{.org}, \texttt{.net}, etc.), along with valuable passive DNS datasets that aid in identifying ccTLD domains. The distinction is that in the 2022 scans, we first identify registered domains through active measurements and then conduct our experiments. However, in the 2017 scans, the data includes both registered and expired domain names from our input list, and we subsequently identify vulnerable registered~domains. 

In total, we sent 1.4 billion update requests for 260M domains and 971K nameserver addresses (both IPv4 and IPv6). The number of successful updates is low. In IPv4 address space, 6,478 successful DNS updates impacted 5,495 domains and 2,072 nameservers. Moreover, a considerable fraction of those (21.4\% of domains and 23.6\% of nameservers) were not fixed since the 2017 global scan. While very few new resources became vulnerable after our notification campaigns, others were insecure for years. Much fewer domains (168) were vulnerable when sending updates to corresponding nameservers over IPv6---73 nameservers being authoritative for 68 domains accepted our updates. 

To gain deeper insights into the popularity of vulnerable domain names, we compiled the Alexa top website lists for the year 2022. The results showed a significant reduction, with only 516 domain names identified as vulnerable, compared to the staggering 5,964 vulnerable domains detected back in 2017---a difference of two orders of magnitude. Moreover, among the 200,000 most popular domain names, we found a mere 172 vulnerable domains. Our evaluation of the notification campaigns presents a positive overall picture, underscoring the effectiveness of our efforts in securing a substantial portion of domains and nameservers from zone poisoning attacks  over the long term.

Finally, in Section \ref{sec:unvealing}, we endeavored to identify indications of zone poisoning attacks in passive DNS and blocklists during 2017. In this section, we propose an alternative methodology and deploy honeypots with a vulnerable configuration to attract potential attackers and detect signs of the attacks in the present day. Specifically, we installed BIND9 as an authoritative nameserver accepting non-secure updates. It hosted 105 domain names, 50 being drop-catch (recently expired) domains appearing in the Alexa top list. We generated full-fledged websites for each hosted domain and obtained TLS certificates. We performed a zone poisoning attack ourselves to ensure it was feasible and managed to take over the domains by updating their \texttt{A} records. Nevertheless, despite attracting numerous port scanners and attackers, the honeypot received no unsolicited DNS \texttt{UPDATE} requests within seven months.

\section{Ethical Considerations\label{sec:ethical}}

Active Internet measurements have become an established practice in computer networking research. The community has developed a set of best practices to guide researchers in conducting measurements ethically~\cite{menlo,zmap}. In our research, we rigorously adhered to the guidelines set forth in the Menlo report~\cite{menlo}, and the ethical considerations outlined by Korczy\'nski et al.~\cite{zonepoisoning} in their preliminary research on zone poisoning. These guidelines and considerations provided us with a solid framework to ensure the ethical conduct~of~our~study.

We created dedicated web pages on each inserted \texttt{researchstudyzp} subdomain, providing clear details about the purpose of our scan and offering contact information. Despite the extremely low probability of a collision with an existing subdomain, we thoroughly assessed this scenario. It should be noted that the newly added record would not overwrite the existing one. All the records were removed after the study. Throughout the process, we received emails from several network operators seeking further details, requesting information about our methodology, and confirming the fixes they had~applied.

Furthermore, we prioritize the principle of beneficence, aiming to benefit the community by identifying vulnerable systems with minimal intervention and taking steps to notify and assist their owners. Our commitment to responsible disclosure involves sharing our findings exclusively with the relevant parties: the authors of the original RFC, CSIRTs, and DNS software vendors, rather than publicly exposing identifiable information about vulnerable resources.

To ensure the highest ethical standards, our research proposal, and accompanying materials, including our approach to conducting active internet scans and notifying CSIRTs, underwent a rigorous review by the human research ethics committee at our institution. The committee carefully evaluated potential risks and concerns, ensuring that appropriate measures were in place to protect the privacy and confidentiality of the domain owners involved.

\section{Related Work\label{sec:related-work}}

Researchers rely on large-scale notifications to inform parties about vulnerabilities (e.g., \cite{remedying,reports,driving,sharing}). However, choosing the appropriate channel, sender, recipient, and framing is challenging. 

Recipient selection is crucial for effective notification. In a study targeting websites with WordPress vulnerabilities, reaching owners directly and indirectly resulted in an overall remediation rate of no more than 26.5\% \cite{problem}. GDPR regulations later made it difficult to retrieve contact emails from domain WHOIS, prompting researchers to find alternative means of reaching domain owners. \texttt{SOA} records were found to be a useful source for administrator emails \cite{wissem}, and CERTs were utilized for dissemination \cite{hell}. Comparing remediation rates between CERT notifications and direct WHOIS notifications showed better performance in the latter group~\cite{li2016you}.

Previous studies focused on non-secure dynamic updates and identified vulnerable domains, but did not perform notifications \cite{zonepoisoning}. In subsequent work, contacting owners via \texttt{SOA}, generic, and WHOIS email addresses resulted in low remediation rates \cite{cetin2017make}. Our research builds upon these studies by introducing an extensive adversary model and conducting Internet-wide measurements, revealing significantly more vulnerable domains \cite{zonepoisoning}. Employing an indirect approach to notifications, we achieved higher remediation rates, with 54\% of nameservers and 98\% of domains being remediated~\cite{cetin2017make}.

\section{Conclusions\label{sec:conclusions}}

In this paper, we analyzed non-secure DNS updates---a standard that lets anyone update the content of DNS zones. We defined an extensive attack taxonomy that shows how a single DNS update packet can enable an attacker to make a domain unavailable, take it over, or compromise the domain control validation. 

We performed a large-scale analysis of more than 354M domains and 3.8M corresponding nameservers. Less than 1\% of vulnerable domains and nameservers, including those of financial, governmental, and healthcare institutions, were identified. We notified national CSIRTs and achieved remediation rates of approximately 98\% for domains and 54\% for nameservers. Our repeated scans in 2022 confirmed the long-term impact of our notifications, with a low population of vulnerable resources. 

Efforts to fix individual systems are labor-intensive and do not provide a guarantee against the recurrence of insecure configurations in the future. Hence, we engaged in discussions with Paul Vixie, the original author of RFC 2136, as well as DNS software vendors, to share the findings of this paper. At the time of writing, two CVEs have been 
reserved for the DNS vendor software, specifically targeting Knot DNS and Simple DNS Plus. 

\section*{Acknowledgment}

The authors express their gratitude to Paul Vixie, CSIRTs, and DNS software vendors for their valuable comments and active participation in the mitigation of vulnerable systems. 
The authors thank the contributors of passive DNS data to Farsight Security and the European Data Sharing Collective (SIE Europe). 
This work has been partially supported by Carnot LSI and Grenoble Alpes Cybersecurity Institute (under the contract ANR-15-IDEX-02), the French Ministry of Research projects PERSYVAL-Lab under contract ANR-11-LABX-0025-01, and DiNS under contract ANR-19-CE25-0009-01. 

\bibliographystyle{plain}
\bibliography{bibliography}

\end{document}